\documentclass{PoS}
\usepackage{epsfig,dcolumn}
\usepackage{graphicx}
\usepackage{color}
\usepackage{amsmath}
\usepackage{amsfonts}
\usepackage{amssymb}
\usepackage{graphicx}
\usepackage{type1cm}
\usepackage{eso-pic}
\usepackage{tabularx}

\newcommand{\be}{\begin{equation}}
\newcommand{\ee}{\end{equation}}

\newcommand{\im}{\mathrm{Im}\,}
\newcommand{\re}{\mathrm{Re}\,}

\newcommand{\degree}{{\rm o}}
\usepackage{xspace}

\newcommand{\gevnospace}{\ensuremath{{\mathrm{\,Ge\kern -0.1em V}}}}
\newcommand{\gev}{\gevnospace\xspace}
\newcommand{\mevnospace}{\ensuremath{{\mathrm{\,Me\kern -0.1em V}}}}

\newcommand{\etapi}{\ensuremath{\eta^{(\prime)}\pi}\xspace}
\newcommand{\pione}
{\ensuremath{\pi_1}\xspace}

\newcommand{\mevp}{\ensuremath{(\!\mevnospace)}}

\usepackage{bm} 

\title{Review of phenomenological analyses of $\etapi$ resonances.}

\ShortTitle{Review of phenomenological analyses of $\etapi$ resonances.}

\author{\speaker{A.~Rodas}\\
        Departamento de F\'isica Te\'orica, Universidad Complutense de Madrid, 28040 Madrid, Spain\\
        E-mail: \email{arodas@ucm.es}}
  \author{A.~Pilloni \\
  Theory Center,
Thomas  Jefferson  National  Accelerator  Facility, 
Newport  News,  VA  23606,  USA \\
        European Centre for Theoretical Studies in Nuclear Physics and Related
Areas (ECT$^*$) and Fondazione Bruno Kessler,
I-38123 Villazzano (TN), Italy\\
        E-mail: \email{pillaus@jlab.org}}
       \author{A.~Szczepaniak \\
       
  Theory Center,
Thomas  Jefferson  National  Accelerator  Facility, 
Newport  News,  VA  23606,  USA \\
        Center for  Exploration  of  Energy  and  Matter,  
Indiana  University,  
Bloomington,  IN  47403,  USA\
Physics  Department,  
Indiana  University,  
Bloomington,  IN  47405,  USA\\
E-mail: \email{aszczepa@indiana.edu}}

\abstract{

We present a robust analysis of the $\etapi$ system in COMPASS data. We fit the extracted relative phases and intensities with a coupled-channel formalism enforcing both unitarity and analyticity.
We provide a robust extraction of a single exotic $\pi_1(1600)$ decaying to both $\etapi$ final states, and the resonance parameters of the $a_2(1320)$ and $a'_2(1700)$. We find no evidence for a second exotic state, which is compatible with recent Lattice QCD estimates. }

\FullConference{XIII Quark Confinement and the Hadron Spectrum - Confinement2018\\
		31 July - 6 August 2018\\
		Maynooth University, Ireland}

\begin{document}

\section{Introduction}

Description of hadron structure in terms of quarks and gluons is key to our understanding of Quantum Chromodynamics (QCD). Although most of the observed mesons can be classified as $q\bar q$ bound states, QCD has a much richer spectrum~\cite{Ketzer:2012vn,Meyer:2015eta,Esposito:2016noz}. 
Several QCD-based models predict states with explicit gluonic degrees of freedom, known as {\em hybrids}~\cite{Horn:1977rq,Isgur:1984bm,Chanowitz:1982qj,Szczepaniak:2001rg,Bass:2001zs}. This predictions have
 been supported by  lattice QCD calculations~\cite{Lacock:1996ny,Bernard:1997ib,Dudek:2013yja}. A single state with quantum numbers $J^{PC} (I^G) = 1^{-+} (1^-)$ below 2 GeV is expected. However, experiments claimed two different states to exist, a $\pi_1(1400)$ decaying into $\eta \pi$, and a $\pi_1(1600)$ decaying into $\rho \pi$ and $\eta'\pi$ channels. The high statistics analyses from COMPASS 
confirmed a 
peak in both $\rho \pi$ and $\eta' \pi$ at around $1.6\gev$~\cite{Alekseev:2009aa,Akhunzyanov:2018pnr} 
and another structure in $\eta \pi$, at $1.4\gev$~\cite{Adolph:2014rpp}. 

In~\cite{Rodas:2018owy} we analyzed the spectrum of the  $\eta \pi$ $D$- and $P$-waves extracted from the COMPASS data with a coupled-channel formalism, extending our previous analysis~\cite{Jackura:2017amb}. We establish the existence of a single $\pione$ in these channels and provide a detailed analysis of its properties.
We also determine the resonance parameters of the $a_2(1320)$ and $a_2'(1700)$.

\begin{figure*}[b]
\centering
\includegraphics[width=.325\textwidth]{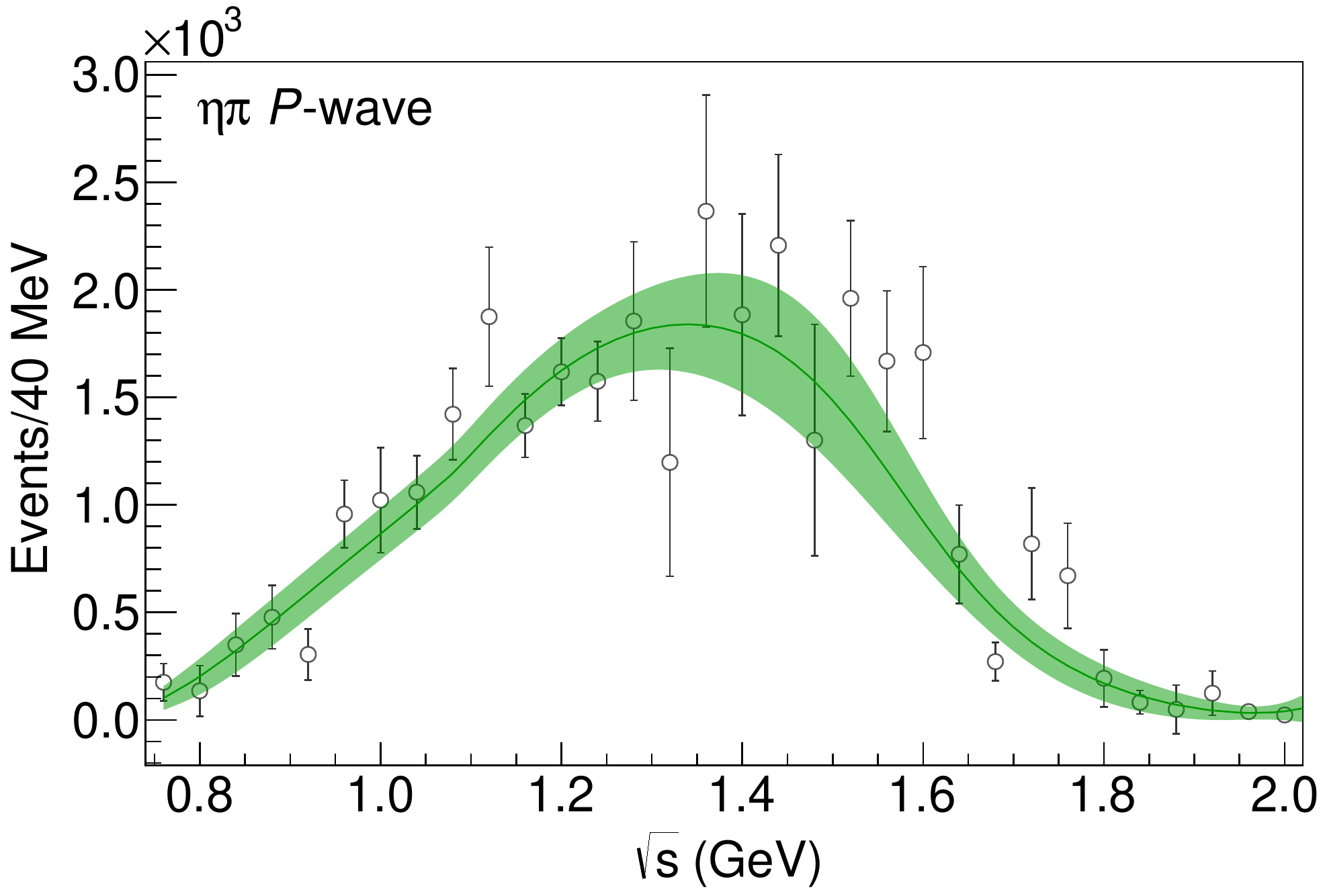}
\includegraphics[width=.325\textwidth]{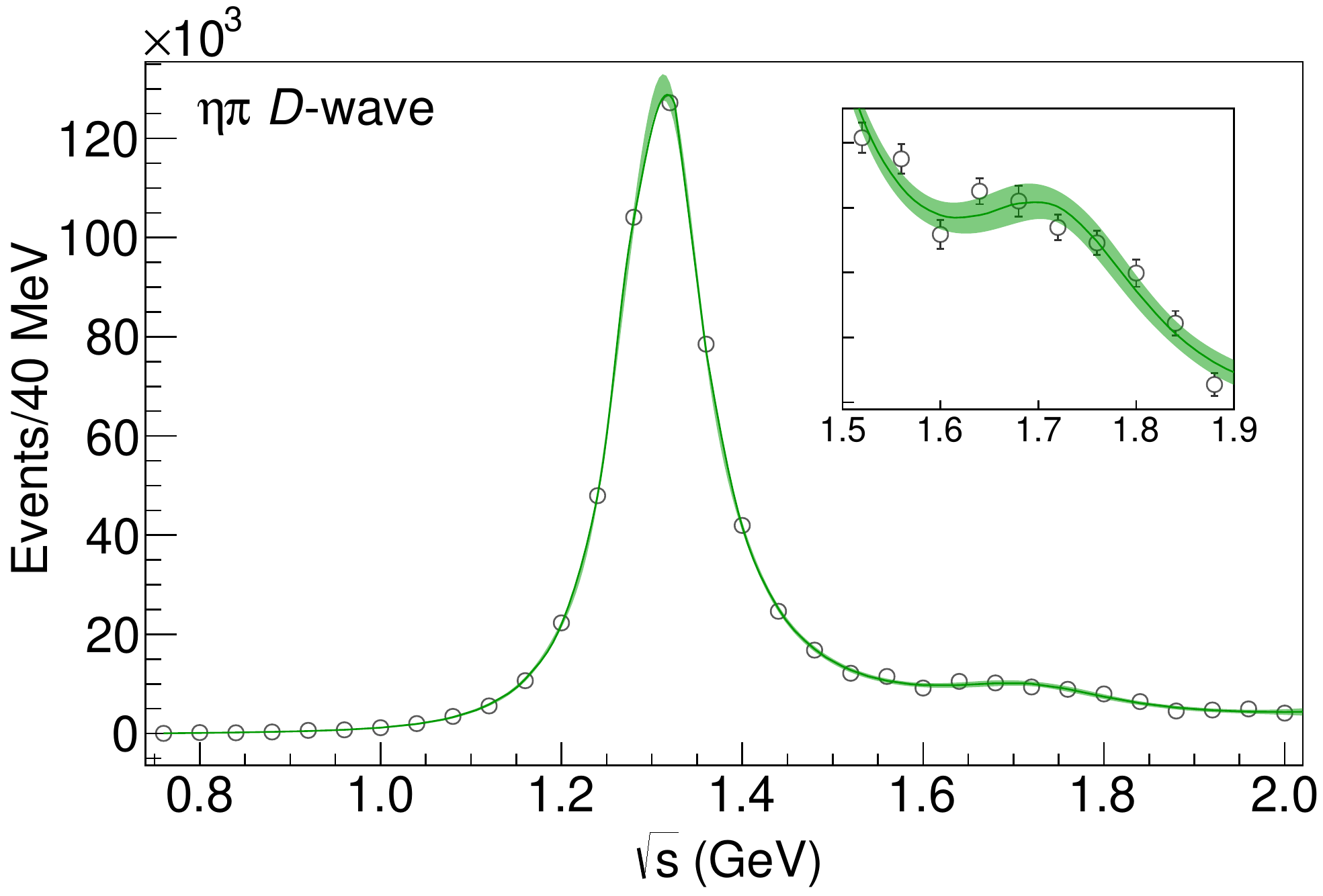}
\includegraphics[width=.325\textwidth]{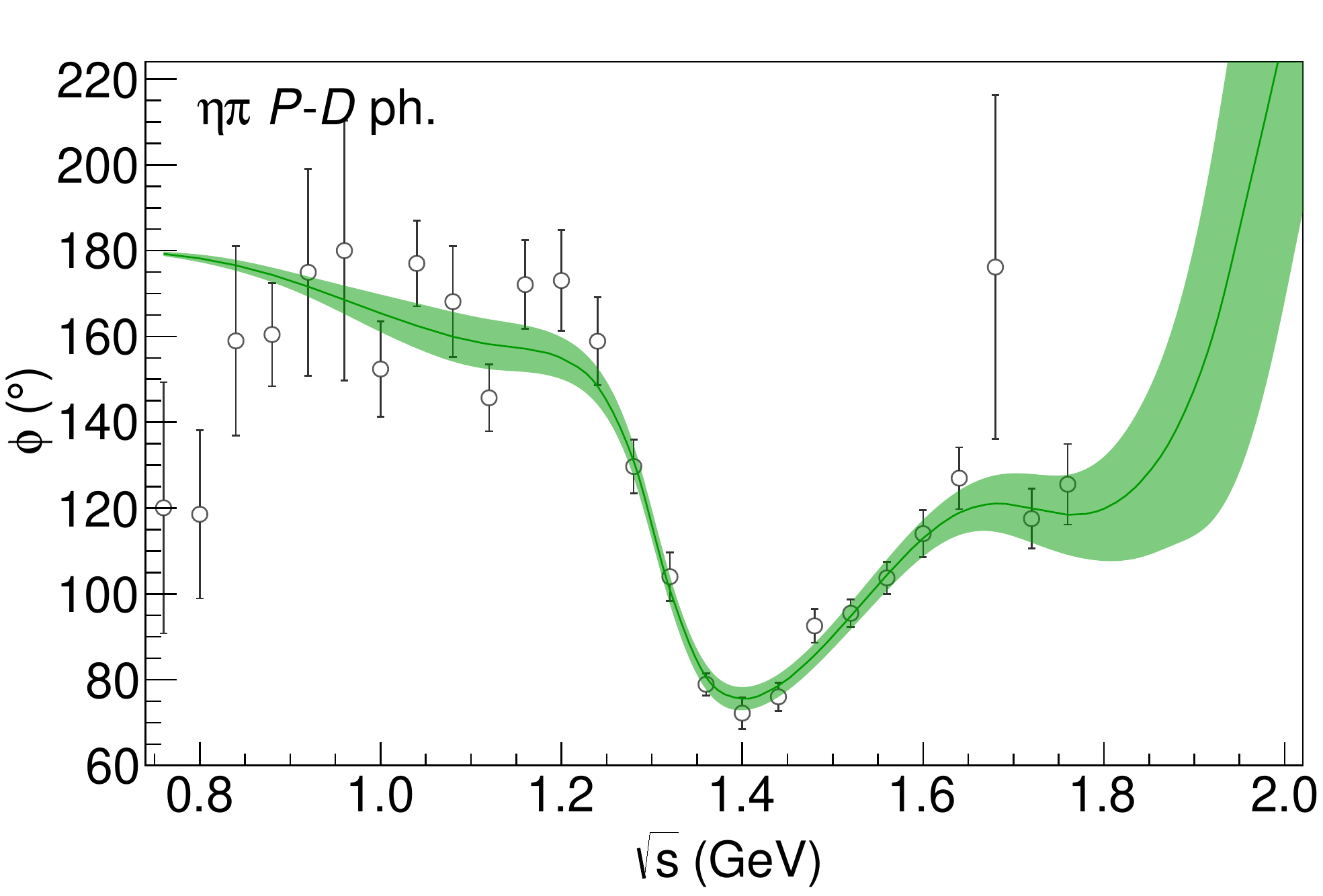}\\
\includegraphics[width=.325\textwidth]{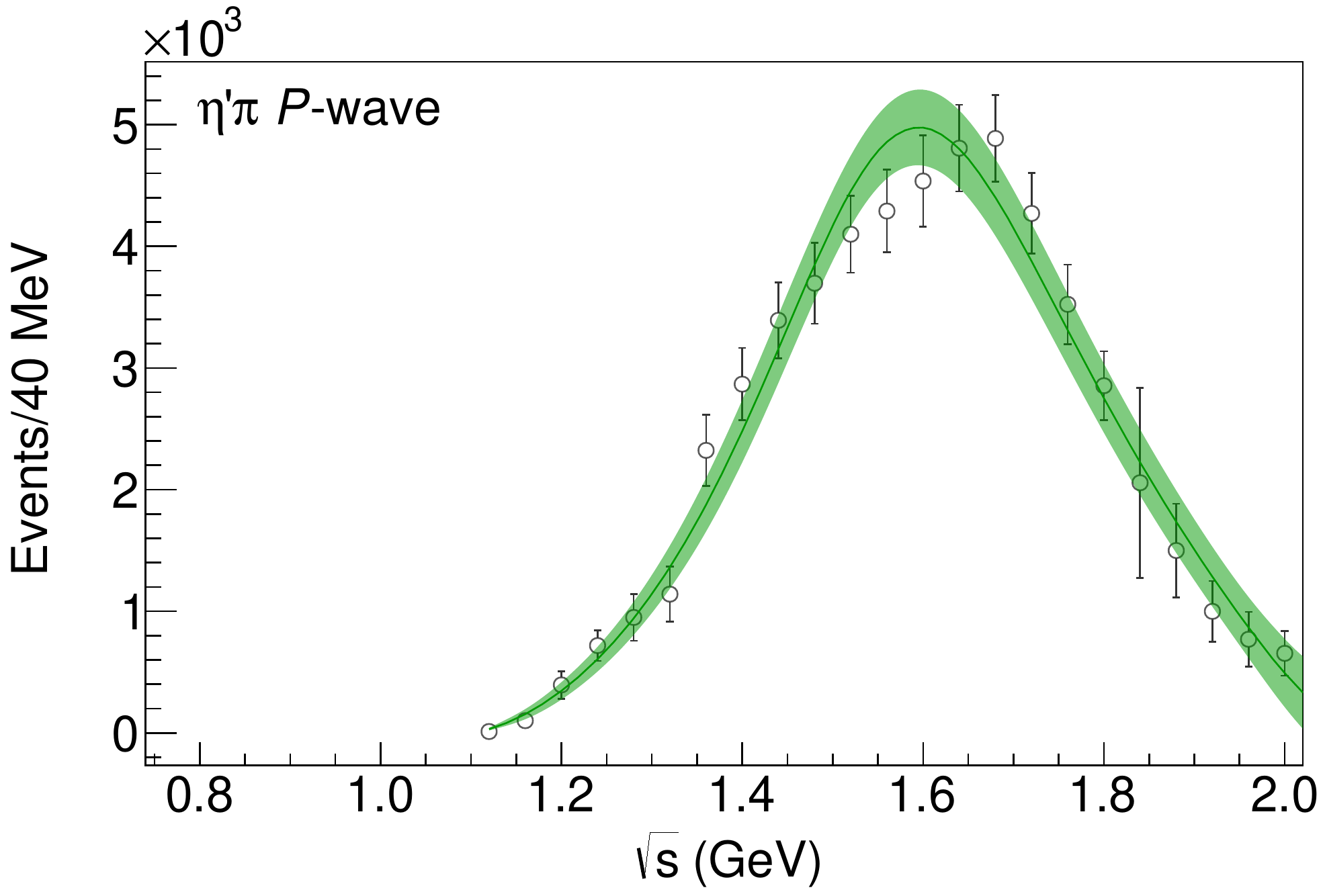}
\includegraphics[width=.325\textwidth]{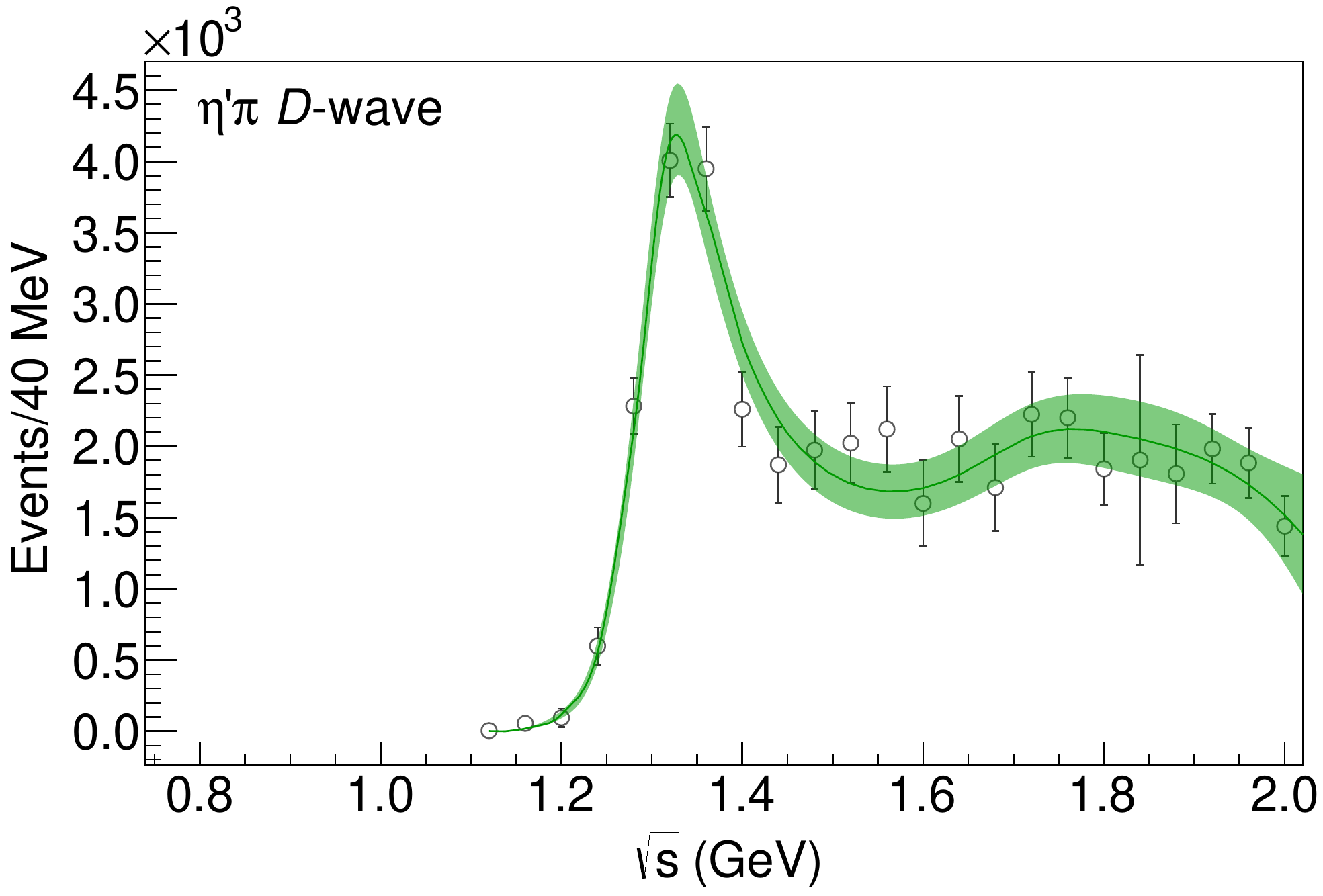}
\includegraphics[width=.325\textwidth]{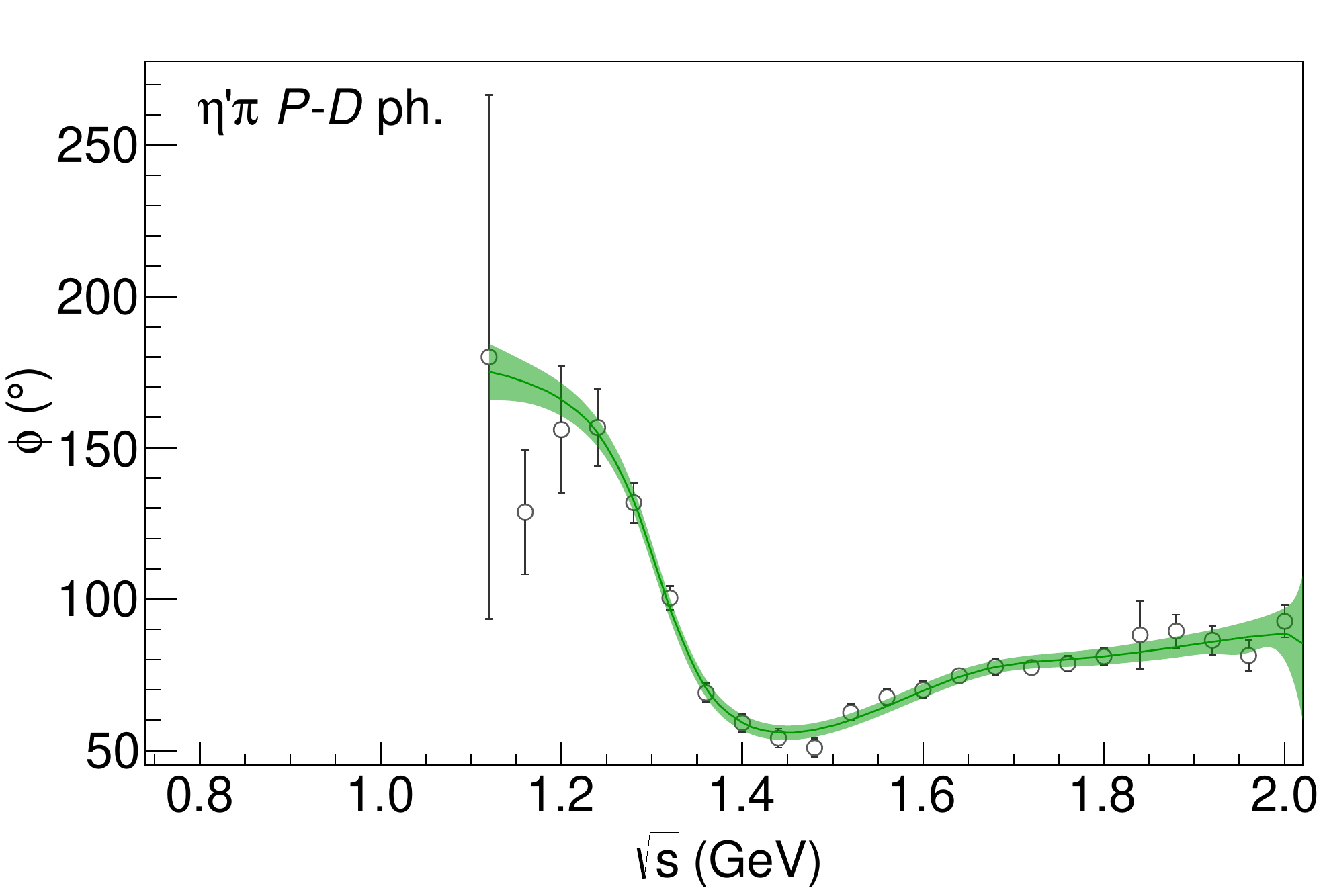}
\caption{From~\cite{Rodas:2018owy}. Fits to the $\eta \pi$ (upper line) and $\eta' \pi$ (lower line) data from COMPASS~\cite{Adolph:2014rpp}.
The intensities of $P$- (left), $D$-wave (center), and their relative phase
(right) are shown. The inset zooms into the region of the $a_2'(1700)$. The solid line and green band shows the result of the fit and the $2\sigma$ confidence level provided by the bootstrap analysis, respectively. The initialization of the fit is chosen by randomly generating $O(10^5)$ different sets of values for the parameters. The best fit has $\chi^2/\text{dof} = 162/122=1.3$. The errors shown are statistical only. 
 } \label{fig:fits}
\end{figure*}

\section{Data}

In our analysis, we focused on the $P$- and $D$-wave partial waves extracted from the COMPASS mass independent analysis of $\pi p\to \etapi p$. Due to the 190~\gev pion beam most of the events are produced in the forward direction, close to the lower limit of the measured transferred momentum squared $-t_1 \in [0.1, 1]\gev^2$. 
In the COMPASS data,  at the  $\eta' \pi$  mass of $2.04\gev$ there is a sharp drop 
 in the $P$-wave intensity,  accompanied by a sudden fall of the phase difference between $P$- and $D$-wave by $50^\degree$. 
 Unfortunately, there exist no data in the $\eta \pi$ channel in the $1.8-2.0\gev$ region, so that we cannot check this behavior. On top of that, fitting these data points of the $P$-wave 
produces nonphysical values for the position of the $a_2'$. For all these 
reasons, we discard the data above $2\gev$.

Recently, COMPASS  published the $3\pi$ partial-wave analysis~\cite{Akhunzyanov:2018pnr}, including the exotic $1^{-+}$ wave in the $\rho\pi$ final state. Unfortunately, the extraction of the resonance pole in this channel is hindered by the irreducible Deck process~\cite{Deck:1964hm,Ascoli:1974hi}. As discussed in~\cite{Jackura:2017amb}, neglecting additional 
channels 
 does not affect the pole position in cases like the one we are studying, so our  analysis will consider only $\etapi$ channels.

\section{Model}

The $\pi p\to \etapi p$ is Pomeron~($\mathbb{P}$) dominated at high energies. This allow us to factorize the $\pi\mathbb{P}\to \etapi$ process, which resembles a helicity partial wave amplitude  $a^J_{i}(s)$ for fixed $t_1$, with $i=\etapi$ the final channel, $J$  the angular momentum of the  \etapi system and $s$ its invariant mass squared. In order to explain the approximately constant hadron cross sections the Pomeron must be spin one, this together with the fact that both angular momentum projections $M=\pm 1$ are related through  parity allow us to drop the Pomeron helicity index. The transferred momentum is fixed to $t_\text{eff}=-0.1\gev^2$.

We parameterize the 
 amplitudes following the coupled-channel $N/D$ 
formalism,
\begin{equation}
\label{eq:amplitude}
 a^J_i(s) = q^{J-1} p_i^J \, \sum_k n^J_k(s) \left[ {D^J(s)}^{-1} \right]_{ki}\,,
\end{equation}
where  
$p_i=\lambda^{1/2}(s,m_{\eta^{(\prime)}}^2,m_{\pi}^2)/(2\sqrt{s})$
is the \etapi breakup momentum, and  
$q=\lambda^{1/2}(s,m_{\pi}^2,t_{\text{eff}})/(2\sqrt{s})$ 
the $\pi$ beam momentum in the \etapi rest frame, with $\lambda(a,b,c)$ being the K\"all\'en triangular function.
The $n^J_k(s)$'s incorporate   exchange ``forces" in the production process (left hand cuts), 
and are smooth functions of $s$ in the physical region. 
The $D^J(s)$ matrix contains the right hand cuts constrained by direct channel unitarity of the $\etapi \to \etapi$ channel interactions.

We use an effective expansion in Chebyshev polynomials for the numerator $n^J_k(s)$.
A customary parameterization of the denominator is given by
\begin{equation}\label{eq:Dsol}
D^J_{ki}(s) =  \left[ {K^J(s)}^{-1}\right]_{ki} - \frac{s}{\pi}\int_{s_{k}}^{\infty}ds'\frac{\rho N^J_{ki}(s') }{s'(s'-s - i\epsilon)}, 
\end{equation}
where
$s_k$ is the threshold in channel $k$ and
\begin{align}
\rho N^J_{ki}(s') &= \delta_{ki} \,\frac{\lambda^{J+1/2}\left(s',m_{\eta^{(\prime)}}^2,m_\pi^2\right)}{\left(s'+s_L\right)^{2J+1+\alpha}} \label{eq:rhoN}
\intertext{ is an effective description of the left
hand cuts in the $\etapi\to\etapi$ scattering, controlled by $s_L$, which is fixed at the hadronic scale $s_L\sim 1\gev^2$. Finally, 
}
K^J_{ki}(s) &= \sum_R \frac{g^{J,R}_k g^{J,R}_i}{m_R^2 - s} + c^J_{ki} + d^J_{ki} \,s,\label{eq:Kmatrix}
\end{align}
with $c^J_{ki}= c^J_{ik}$ and $d^J_{ki}= d^J_{ik}$,
is a standard parameterization for the $K$-matrix.
We consider two $K$-matrix poles in the $D$-wave, and one single $K$-matrix pole in the $P$-wave when obtaining our best fit to data; 
the numerator of each channel and wave is described by a third-order polynomial, and
we set $\alpha = 2$ in Eq.~\eqref{eq:rhoN}. The remaining 37 parameters are fitted to data.
 The best fit has $\chi^2 / \text{dof} = 162/122 = 1.3$, 
 in good agreement with data as shown in Fig.~\ref{fig:fits}. In particular, a single $K$-matrix pole is able to correctly describe the $P$-wave peaks in the two channels.
The uncertainties on the parameters have been estimated via the bootstrap method. 

\begin{figure*}
\includegraphics[width=\textwidth]{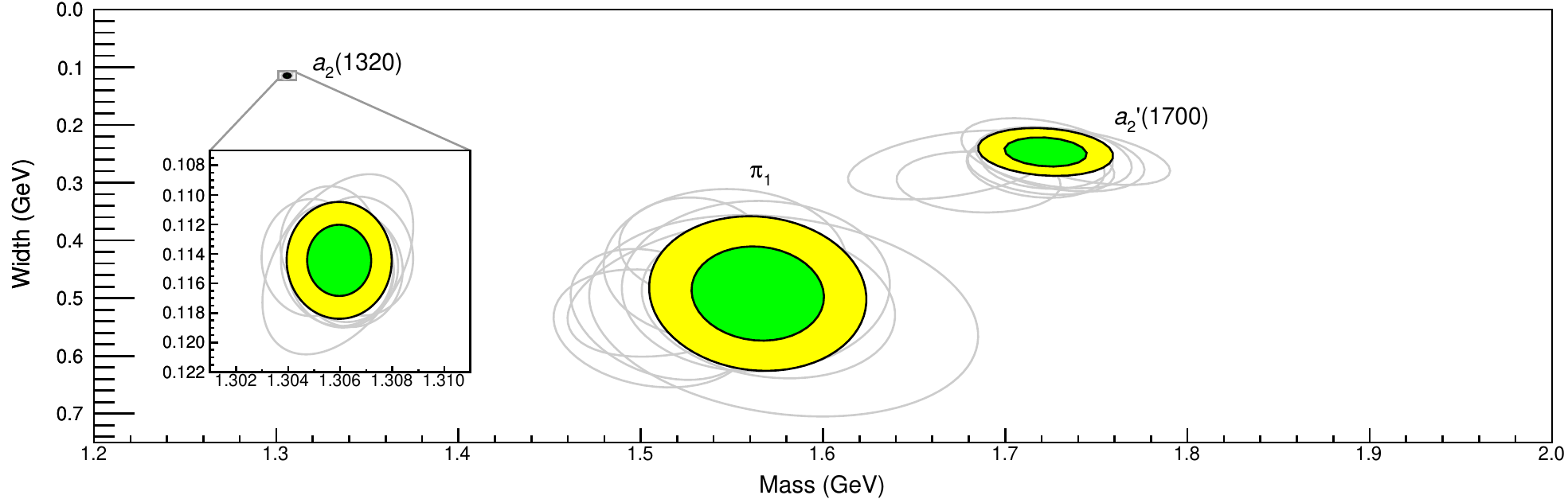}
\caption{\label{fig:poles} From~\cite{Rodas:2018owy}. Positions of the poles identified as the $a_2(1320)$, $\pione$, and $a_2'(1700)$. The inset shows the position of the $a_2(1320)$. The green and yellow ellipses show the $1\sigma$ and $2\sigma$ confidence levels, respectively. The gray ellipses in the background show, 
 within $2\sigma$, variation of the pole position  upon changing the functional form and the parameters of the model, as discussed in the text
 }
\end{figure*}

Once the fits are obtained, the $D^J(s)$ matrix in Eq.~\eqref{eq:Dsol} can be continued underneath 
the unitarity cut into the closest Riemann sheet. A pole $s_P$ in the amplitude appears when the determinant of $D^J(s_P)$ vanishes. The poles close to the real axis drive the behavior of the partial waves in the real axis, these can be identified as resonances.
In a coupled-channel problem, it is not possible to  specify the number of poles. Appearance of spurious poles far from the physical region is likely. However one could isolate the physical poles by testing their stability against different parameterizations and data resampling. 
We select the resonance poles in the $m\in [1,2]\gev$ and $\Gamma \in [0,1]\gev$ region, where customarily $m = \re \sqrt{s_P}$ and $\Gamma = -2\im \sqrt{s_P}$.
Two poles are found in the $D$-wave, identified as the $a_2(1320)$ and $a_2'(1700)$, and a single pole in the $P$-wave, which we call $\pione$. The pole positions are shown in Fig.~\ref{fig:poles}, while the resonance parameters are listed in Table~\ref{tab:poles}. 
We have also performed a pure background fit for $J=1$, obtaining a $\chi^2$ larger by almost two orders of magnitude when no pole is found, thus rejecting the possibility for the $P$-wave peaks to be generated by non-resonant production.

Regarding the existence of two different states we have considered solutions with two isolated $P$-wave  poles, generated by using more K-matrix poles. This is the scenario discussed 
 in the PDG, and although the $\chi^2$ for this case is equivalent to the reference fit, one of the poles can appear in a large region depending on the initial values of the fit, while the second one is compatible with the single pole solution. The former does not influence the real axis close to its position but changes the behavior of the phase, now having a $180^\degree$ jump where no data exist. We thus conclude it is just an artifact of including a second pole having no physical meaning.

\section{Systematic uncertainties}
The pole extraction requires an analytic model which carries systematic uncertainties. Regarding the numerator, which is expected to be smooth, we have varied $t_\text{eff}$ and the order of the polynomial. As for the denominator, we have first varied the values of $s_L$ and $\alpha$ in a considerable range. Finally we have also modified the Chew-Mandelstam term, to include the phenomenological description of a $t$-channel exchange dominated by an intermediate particle, which mass is considered to be of the order of 1\gev, explicitly the term reads
\begin{equation}
\rho N^J_{ki}(s') = \delta_{ki}\, Q_J(z_{s'})\, s'^{-\alpha} \lambda^{-1/2} (s^{\prime},m^2_{\eta^{(\prime)}},m^2_\pi) ,
\end{equation}
where $Q_J(z_s)$ is the second kind Legendre function, and $z_{s'}=1+2s' s_L/\lambda (s^{\prime},m^2_{\eta^{(\prime)}},m^2_\pi)$  the scattering angle of the elastic scattering, and $s_L = 1\gev^2$.  This function behaves asymptotically as $s^{-\alpha}$, has a left hand cut starting at $s = 0$, a short cut between $(s^{\prime}-m_{\eta^{(\prime)}})^2$ and $(s^{\prime}+m_{\eta^{(\prime)}})^2$, and an incomplete circular cut.

The shape of the dispersive integral in Eq.~\eqref{eq:Dsol} is altered, but the fit quality is unaffected under all these changes. The pole positions change roughly within $2\sigma$, as shown Fig.~\ref{fig:poles}, while systematic uncertainties are reported in Table~\ref{tab:poles}.

\begin{table}
\caption{Resonance parameters. The first error is statistical, the second systematic.}
\begin{center}
\begin{tabular}{c c c} 
\hline 
\hline
Poles & Mass \mevp & Width \mevp \\ \hline 
$a_2(1320)$ & $1306.0 \pm 0.8 \pm 1.3 $ & $114.4 \pm 1.6 \pm 0.0$ \\ 
$a_2'(1700)$ & $1722 \pm 15 \pm 67 $ & $247 \pm 17 \pm 63$ \\ 
$\pione$ & $1564 \pm 24 \pm 86 $ & $492 \pm 54 \pm 102$ \\ 
\hline
\end{tabular}
\end{center}
\label{tab:poles}
\end{table}

\section{Summary}

We used a standard $K$-matrix formula constrained by unitarity and analiticity~\cite{Rodas:2018owy} to perform the first coupled-channel analysis in the $\etapi$ system measured at COMPASS~\cite{Adolph:2014rpp}. Two ordinary mesons, identifies as the $a_2(1320)$ and the $a_2'(1700)$ are found in the $D$-wave. In the $P$-wave however, a single exotic pole $\pione$ is obtained, compatible with the Lattice QCD~\cite{Lacock:1996ny,Bernard:1997ib,Dudek:2013yja} suggestion of a single isovector with $J^{PC}=1^{-+}$ quantum numbers. Its mass and width are determined to be $1564\pm 24\pm 86$ MeV and $492\pm 54\pm 102$ MeV, respectively. The systematic uncertainties are determined through the variation of both parameters and functional forms that are not directly constrained. There is no evidence of the existence of a second exotic state.

\vspace{-0.2cm}
\section{Acknowledgments}
This work was supported by
the U.S.~Department of Energy under grants
No.~DE-AC05-06OR23177 
 and No.~DE-FG02-87ER40365, 
U.S.~National Science Foundation under award number
PHY-1415459, 
 and Ministerio de Ciencia, Innovaci\'on y Universidades (Spain)
grant FPA2016-75654-C2-2-P.
AR acknowledges the Universidad Complutense for a doctoral fellowship.
\vspace{-0.25cm}

\bibliographystyle{JHEP}
\bibliography{quattro.bib}

\end{document}